\documentclass{osa-article}

\journal{oe}

\usepackage{graphicx} 

\usepackage{subcaption} 
\usepackage{dcolumn}
\usepackage{bm}
\usepackage{booktabs}
\usepackage{gensymb}
\usepackage{adjustbox}
\usepackage{amsmath}
\usepackage{siunitx}

\usepackage[english]{babel}

\begin{document}

\title{Long-term hybrid stabilization of the carrier-envelope phase}

\author{Jack Hirschman,\authormark{1,2,*} Randy Lemons,\authormark{2,3} Evan Chansky, \authormark{2,4} G{\"u}nter Steinmeyer,  \authormark{5} and Sergio Carbajo \authormark{2,3}}

\address{\authormark{1}Department of Applied Physics, Stanford University, 348 Via Pueblo, Stanford, CA 94305, USA\\
\authormark{2}SLAC National Accelerator Laboratory, 2575 Sand Hill Rd, Menlo Park, CA 94025, USA\\
\authormark{3}Colorado School of Mines, 1500 Illinois St, Golden, CO 80401, USA\\
\authormark{4}Lehigh University, 27 Memorial Drive W, Bethlehem, PA 18015, USA\\
\authormark{5}Max Born Institute for Nonlinear Optics and Short Pulse Spectroscopy, Max-Born-Stra{\ss}e 2a, 12489 Berlin, Germany}

\email{\authormark{*}jhirschm@stanford.edu} 



\begin{abstract*}
Controlling the carrier envelope phase (CEP) in mode-locked lasers over practically long timescales is crucial for real-world applications in ultrafast optics and precision metrology. We present a hybrid solution that combines a feed-forward technique to stabilize the phase offset in fast timescales and a feedback technique that addresses slowly varying sources of interference and locking bandwidth limitations associated with gain media with long upper-state lifetimes. We experimentally realize the hybrid stabilization system in an Er:Yb:glass mode-locked laser and demonstrate 75 hours of stabilization with integrated phase noise of \SI{14}{\milli\radian} (1 \SI{}{\hertz} to 3 \SI{}{\mega\hertz}), corresponding to around \SI{11}{\atto\s} of carrier to envelope jitter. Additionally, we examine the impact of environmental factors, such as humidity and pressure, on the long-term stability and performance of the system.
\end{abstract*}

\section{\label{sec:intro}Introduction}
 With lasers operating in the femtosecond and attosecond regimes becoming increasingly prevalent, precisely controlling the offset of the underlying electric field with respect to the envelope of the pulse, known as the carrier envelope offset phase (CEP), will continue to rise in importance. Myriads of applications including optical frequency metrology~\cite{Udem2002, Krausz2014}, high harmonic generation~\cite{Sansone443, Sola2006}, coherent synthesis of distributed fiber-laser arrays~\cite{Ye2002, Lemons2020}, and laser-based~\cite{Salamin2019, Carbajo2016, Wong2017} and on-chip accelerators~\cite{Sapra79, Cesar:18} rely on stabilizing and controlling the CEP. Achieving this stabilization and control over practically long durations is just as important if it is to be used in real-world applications. 

There are two common CEP locking schemes in mode-locked lasers, feedback (FB) and feed-forward (FF). FB techniques measure fluctuations in the carrier-envelope offset (CEO) frequency and make adjustments to the laser cavity or to the pump power in order to alter differences in the phase and group velocities~\cite{Jones:2000, Yu:07, Udem:99}. Specifically, changes to the cavity directly affect the path length and changes to pump power alter the phase and group velocities via nonlinear intracavity processes. These FB methods require electronics, usually in the form of proportional-integral-derivative (PID) controllers or phase-locked loops, designed to maintain the locked CEO frequency. 

FF  techniques instead allow for phase changes in the cavity to develop naturally and then provide a means for modulating the pulse phase downstream~\cite{Lucking:12, Koke:2010, Steinmeyer2013}. Here, CEP can be stabilized and controlled using an acousto-optic frequency shifter (AOFS). Furthermore, FF techniques often require relatively simple electronics to achieve stabilization and do not tie short-term phase stabilization to longer-term system performance. However, phase modulators like an AOFS have limited operation bandwidths (BW), typically on the order of a few hundred kHz, which can compromise long-term stabilization if the modulating frequency drifts away from the BW range.

Prior work has explored retaining stability using FF techniques over runs as long as a day. One of the longest runs came from L\"{u}cking, et al., who report less than \SI{30}{\milli\radian} of phase noise with a CEP-locked run of more than 24 hours~\cite{Lucking:12} using a Ti:sapphire oscillator. Zhang, et al., using a Kerr-lens mode-locked Yb-CYA \SI{1}{\micro\m} solid-state laser, report a run stabilized over two hours with a residual integrated phase noise of \SI{79.3}{\milli\radian} (\SI{1}{\Hz} to \SI{1}{\mega\Hz})~\cite{Zhang:19} using purely FF techniques. For solid-state lasers in the \SI{1}{\micro\m} range, slightly longer runs can be achieved using FF methods by manually adjusting certain system parameters. For instance, Lemons, et al., using an Er:Yb:glass laser oscillator, could achieve eight hours of stabilization using manual adjustments to the pump power during run time but introducing large amounts of low-frequency phase noise as a result~\cite{Lemons:19}. Replacing the manual adjustments with adequate FB enables longer stabilization durations. Recently, utilizing a mix of FF and FB techniques, Musheghyan, et al. presented 75 hours of locked CEP in a Ti:Sapphire amplifier with integrated phase noise of \SI{150}{\milli\radian}~\cite{Musheghyan:19}. 

As one possible solution to achieve ultra-low noise CEP stabilization over practically long periods of time, in this letter, we elaborate on prior work to examine a FF technique combined with a slow-drift compensation FB system on a Er:Yb:glass mode-locked laser which does not detrimentally affect the short-term phase noise performance. After providing further motivation for combining FF and FB, we will present an overview of the CEP stabilization system and will describe in detail the slow-drift compensation FB system design and performance, which is susceptible to environmental variations, such as temperature, humidity, and pressure changes.

\begin{figure*}[htbp]
	\centering
	\includegraphics[width=1\textwidth]{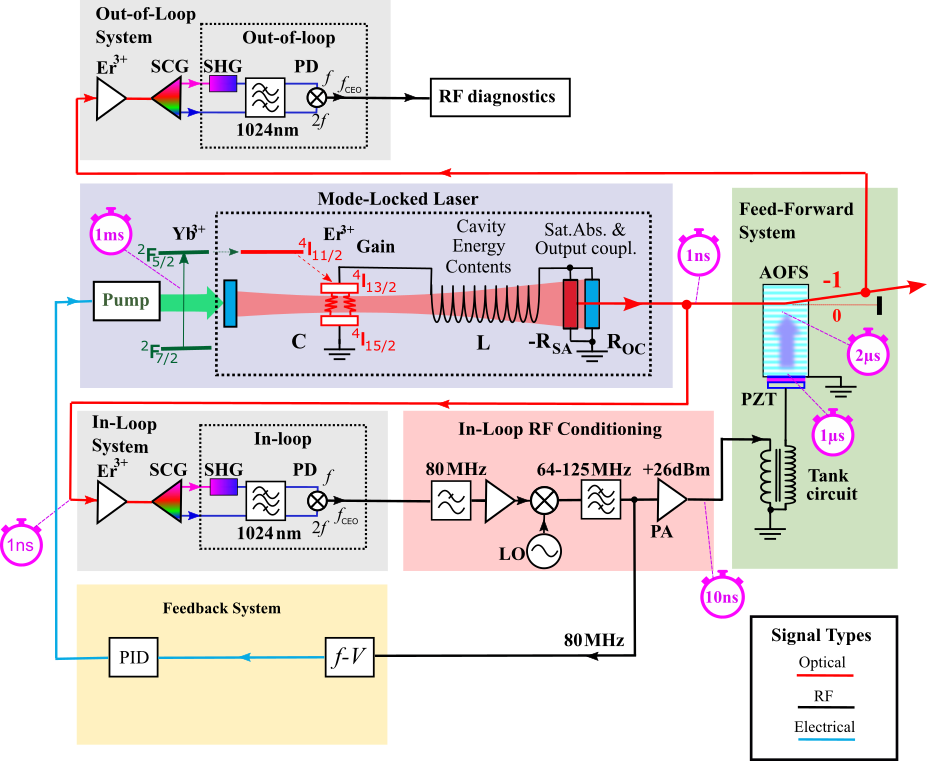}
	\caption{The system contains seven functional blocks: Mode-locked laser, IL $f$-to-$2f$, IL RF conditioning, OOL $f$-to-$2f$, OOL RF conditioning, FF system, and FB system. Signals in red are optical , signals in black are RF paths, and signals in light blue are electrical. }
   \label{fig:detailedSystemOverview}
\end{figure*}

\section{\label{sec:ffFbMotivation}Motivation for a hybrid  feed-forward and feedback approach}
CEP stabilization in Ti:Sapphire lasers and certain fiber lasers has become increasingly common with some commercial systems available today. On the other hand, stabilization of all-solid-state lasers in the 1 to 2 \SI{}{\micro\m} wavelength range has proven more difficult to accomplish. An important factor to consider is the relatively long upper-state lifetime of Yb$^{3+}$ ranging around 1 to 2 \SI{}{\milli\s} for the $^2$F$_{5/2}$ to $^2$F$_{7/2}$ transition~\cite{Gan1995} or that of Er:doped host materials, typically around 8 to 10 ms for $^4$I$_{13/2}$~\cite{Digonnet2001}. But considering that  Er:fiber lasers rely on the same ionic transition in Er$^{3+}$ as the Er:Yb:glass laser in this study motivates examining the interplay between these time constants and their role in gain dynamics of mode-locked lasers in careful detail. For this exercise, we employ a textbook model~\cite{yariv1989, morishita1979, katz1981, tucker1981}, which describes the gain dynamics as an effective resonant RLC circuit. In this model, the population inversion in the gain medium is equivalent to the energy stored in the capacitor C, the intracavity pulse energy translates into magnetic energy stored in an inductor L, and output coupling (OC) along with other intracavity losses are expressed as a resistance R$_{\rm OC}$. Additionally, for this class of mode-locked lasers, the model requires a second resistance for saturated absorption (SA) represented as R$_{\rm SA}$, which has a negative effective value and is in parallel with R$_{\rm OC}$.

Most mode-locked fiber lasers rely on an instantaneous reactive nonlinearity, e.g., a nonlinear optical loop mirror~\cite{Doran:88} whereas all-solid-state lasers are typically mode-locked by soliton mode-locking~\cite{kartner1996} with a semiconductor saturable absorber. While this method is widespread and very successfully commercialized, it strongly relies on suitable suppression of q-switching instabilities~\cite{Honninger:99}. To this end, the relative spot size in the gain medium and on the absorber have to be carefully adapted to avoid positive feedback within the effective RLC circuit. For sustained mode-locking, the nonlinear feedback cannot be arbitrarily reduced, and, as a result, the gain dynamics of mode-locked all-solid-state lasers are either critically damped or even slightly underdamped. 

Limitations in active feedback stabilization by the long upper-state lifetime of Er:Yb:glass can be overcome by feeding back to a fast intracavity absorber rather than the pump power~\cite{Hoffmann:13, Lee2016}. An alternative stabilization method that is capable of circumventing the upper-state lifetime and locking bandwidth constraints is the feed-forward scheme, which employs an acousto-optic frequency shifter to correct the CEP outside the laser cavity~\cite{Koke:2010}. Specifically, the AOFS modulates the phase of the pulses once they are coupled out of the mode-locked laser. Using the first order diffracted beam from the AOFS, the frequency comb of the laser is altered and the CEO frequency becomes effectively locked~\cite{Lemons:19, AOFS}. This has proven to be an elegant solution on short timescales, resulting in residual jitters as low as a few milliradians~\cite{Lemons:19}.

 On longer timescales, however, the FF scheme exhibits significant phase drift, the origin of which manifests in analyzing the signal path from the laser to the AOFS, and ultimately results in a limited operation bandwidth for the AOFS. Slow drifts can alter $f_{\rm CEO}$ to the extent that the RF drive frequency deviates so far from the center frequency of the AOFS that the Bragg angle is significantly altered~\cite{Steinmeyer2013}, thus negatively impacting the diffracted efficiency. This phenomenon can be understood by analyzing timing within a FF setup, as depicted in Figure~\ref{fig:detailedSystemOverview}. Travel times from the laser to radio frequency (RF) output of the $f$-to-$2f$ interferometer are generally kept in the few nanosecond range. Comparatively large group delays occur in the AOFS, which relies on the generation of an ultrasonic wave with a typical travel velocity on the order of $10^{-5}$ of the vacuum speed of light. For example, bringing the input beam to the AOFS as close as a few hundred microns to the piezoelectric transducer on the AOFS still generates a signal delay greater than \SI{1}{\micro\s}. Furthermore, a similar delay is generated by resonant circuitry in the AOFS used to convert the output current of a microwave amplifier into high voltage amplitude necessary for driving the transducer for a high diffraction efficiency of the AOFS.

The lag in the signal path to the AOFS does not pose a problem when used at constant frequency as resulting phase shifts can be compensated. However, with the CEP freely drifting inside the laser, an error in CEP correction can occur at the AOFS. With an assumed \SI{2}{\micro\s} as an optimistic assumption, a drift in frequency of \SI{100}{\kilo\Hz} translates to a drift in phase of over \SI{1}{\radian}. In this paper, we present a feedback scheme to enable uninterrupted long-term operation of feed-forward CEP stabilization which can overcome the bandwidth constraints from an exclusively feed-forward approach~\cite{Borchers:11} (as depicted in Figure~\ref{fig:detailedSystemOverview} and more thoroughly described in Section~\ref{sec:systemOverview}). FF stabilization ensures high CEP stability down to millisecond timescales, and then at longer durations, a simple frequency stabilization takes over to keep the drive frequency of the AOFS near its center frequency.

\section{\label{sec:systemOverview} System overview}
The system achieves CEP stabilization via FF short-term stabilization combined with a slow-drift FB for stabilization over many hours. The detailed schematic diagram is shown in Figure~\ref{fig:detailedSystemOverview}. There are seven main functional blocks: the mode-locked laser (an Origami-15 Er:Yb:glass laser), the in-loop (IL) $f$-to-$2f$ interferometer, the IL radio frequency (RF) conditioning, the FF AOFS system, the AOFS-based feed-forward system, the PID-based FB system, the out-of-loop (OOL) $f$-to-$2f$ interferometer, and the OOL RF diagnostics. Within the system there is one main signal path with two auxiliary paths (FF and FB). The main signal path is predominately optical. It is generated in the laser cavity, feeds into the AOFS, goes through the OOL $f$-to-$2f$ interferometer, and is then converted into RF for diagnostics. The feed-forward auxiliary path is a mix of optical and RF. It starts with an optical output from the laser cavity, goes through the IL $f$-to-$2f$ interferometer where it is converted into an RF signal, and then enters the IL RF conditioning section where it feeds into the AOFS and can affect the signal path. The feedback auxiliary path is a mix of optical, RF, and electrical signals. It at first follows the FF auxiliary path but, instead of going into the AOFS, branches-off and feeds into a PID controller which modulates pump power to the cavity.

The IL and OOL segments operate in nearly identical fashion, utilizing the $f$-to-$2f$ heterodyning technique. For the path through the IL System, the laser spectrum can be quantified as 
\begin{equation}
    \centering
    \label{modeLockedComb}
    f_n = n f_{\rm REP} + f_{\rm CEO},
\end{equation}
where $f_n$ is the comb spacing of the mode-locked laser and $f_{\rm REP}$ is the repetition rate of the mode-locked laser. This optical comb from the original spectrum is amplified and compressed in order to reach intensities necessary for non-linear broadening of the frequency spectrum to ensure $f_{2n}$ is reached (seen in Figure~\ref{fig:detailedSystemOverview} as SCG, super continuum generation). The pulse then undergoes frequency doubling to obtain $2f_n$ (seen in Figure~\ref{fig:detailedSystemOverview} as SHG, second harmonic generation). Filters select these two signals at the same frequency, and the heterodyning occurs at the photodiode (PD) where the signal is output as an RF signal. This part of the system is described in detail in Ref.~\cite{Lemons:19}.

The RF output signal from IL now contains the $f_{2n}$ and $2f_n$ mixing products and is fed into the IL RF conditioning segment before feeding into the AOFS and the FB loop. The RF conditioning is required because the operating range of the AOFS is $80 \pm 2.5$ \SI{}{\mega\Hz}, and $f_{\rm CEO}$ may not be in this range. The first set of components in the IL RF conditioning use low pass filters to retain the difference mixing term: 
\begin{equation}
\centering
\label{eq:fCEOExtract}
\begin{split}
    &2f_n - f_{2n} \\&= 2nf_{\rm REP} + 2f_{\rm CEO} - ( 2nf_{\rm REP} + f_{\rm CEO}) = f_{\rm CEO}.
\end{split}
\end{equation}
The $f_{\rm CEO}$ signal is amplified and obtains a signal-to-noise ratio (SNR) of \SI{40}{\deci\bel} (with resolution bandwidth: \SI{100}{\kilo\hertz}) before being mixed with a local oscillator (LO) signal to ensure the output signal is in the correct band for the AOFS.

The LO signal comes from a RF frequency comb generated from an ultra-low phase noise, \SI{10}{\mega\Hz} rubidium crystal oscillator (Stanford Research Systems PRS10) signal fed into a divider that creates a comb spacing of \SI{1.465}{\mega\Hz} starting from \SI{1.4}{\mega\Hz}. The LO signal used is only one line of this comb. It is chosen for mixing with $f_{\rm CEO}$ via a band-pass (BP) filter such that 
\begin{equation}
    \centering
    \label{eq:fAOFS}
    f_{\rm AOFS} = f_{\rm CEO} + f_{\rm LO} = 80\, \text{MHz}.
\end{equation}
The mixed $f_{\rm AOFS}$ signal then follows two branches: one into the AOFS for FF, the other into the FB system. Using the $-1^{st}$ diffraction order of the AOFS yields
\begin{equation}
\centering
\label{eq:OOL}
\begin{split}
    &f_{\rm OOL} \\&= f_n - f_{\rm AOFS} \\&= (n f_{\rm REP} + f_{\rm CEO}) - (f_{\rm CEO} + f_{\rm LO})
    \\&= n f_{\rm REP} - f_{\rm LO},
\end{split}
\end{equation}
where negative and positive frequencies are physically equivalent. This output is sent into the OOL segment where the signal goes through a similar process as in the IL segment. The output from the OOL PD has the mixed $f_{\rm REP}$ and $f_{\rm LO}$ products as well as the constituents. The RF output of the OOL PD then undergoes low pass filtering in the OOL RF Diagnostics block, yielding a measurement of $f_{\rm LO}$.

The repetition rate of the mode-locked laser, $f_{\rm REP}$, is \SI{204}{\mega\Hz}, and $f_{\rm LO}$ was chosen in order to keep $f_{\rm AOFS}$ at \SI{80}{\mega\Hz}. Since $f_{\rm CEO}$ and $f_{\rm LO}$ are related via $f_{\rm AOFS}$ (see equation~\ref{eq:fAOFS}), then the stability of $f_{\rm CEO}$ can be determined from the stability of $f_{\rm LO}$ and the stability of $f_{\rm AOFS}$, and in turn one can monitor drifts in $f_{\rm OOL}$ and $f_{\rm AOFS}$ as a means for characterizing how stable the system is (see equation~\ref{eq:OOL}).

The other branch out of the IL RF conditioning goes into the FB block. Section~\ref{sec:FB} discusses this system in detail.

\section{\label{sec:FB}Slow drift FB stabilization}
\begin{figure*}[htbp]
    \centering
    \includegraphics[width=1\textwidth ]{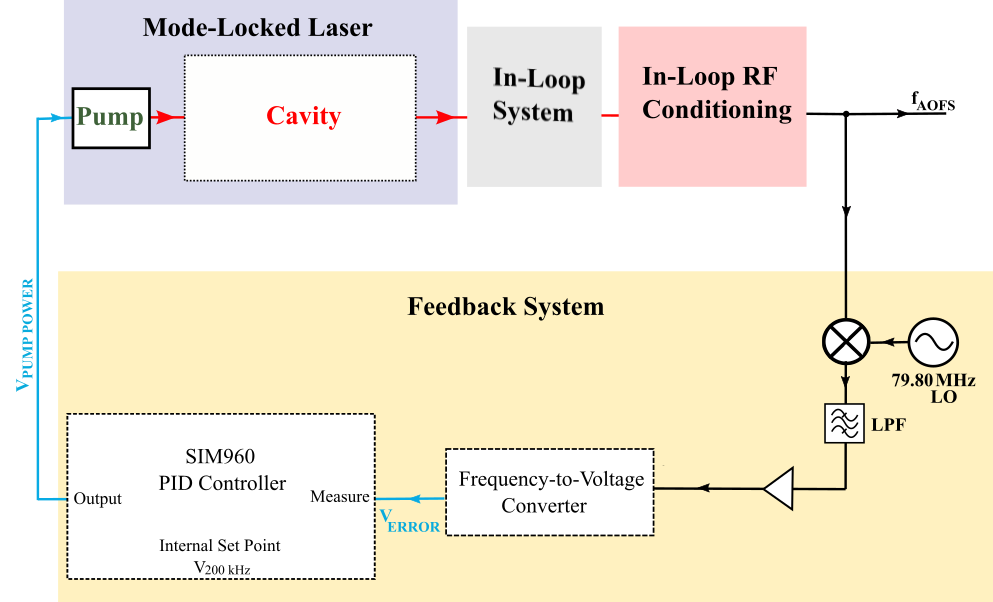}
    \caption{The FB system consists predominantly of RF conditioning, which generates an error frequency from the input $f_{\rm AOFS}$ signal, and a FB control system, which converts this error frequency to a voltage and produces a voltage to drive the pump power in order to lock $f_{\rm AOFS}$ at a particular frequency.}
    \label{fig:FBSystem}
\end{figure*}

The FB system works by generating an error signal from the $f_{\rm AOFS}$ signal and feeding this error signal to a PID controller, which in turn affects the pump power to the cavity. Figure~\ref{fig:FBSystem} shows how this FB is accomplished.

The $f_{\rm AOFS}$ signal is produced by the IL RF conditioning block, as discussed in Section~\ref{sec:systemOverview}. It is also the sum of $f_{\rm CEO}$ and the constant $f_{\rm LO}$ and is set for optimal operation at \SI{80}{\mega\Hz}. Therefore, the deviation in $f_{\rm AOFS}$ away from \SI{80}{\mega\Hz} can be used as the error signal. To generate the error signal from these deviations, $f_{\rm AOFS}$ is mixed with a local oscillator ($f_{{\rm SET}}$) of \SI{79.80}{\mega\Hz}, whose filtered mixing products would be $f_{\rm ERROR}$: $f_{\rm ERROR} = f_{\rm AOFS} - f_{{\rm SET}}$. The operation is limited by the frequency-to-voltage converter chip with input frequency range of 0 to 500 \SI{}{\kilo\Hz}, where $f_{\rm AOFS}$ between \SI{79.80}{\mega\Hz} and \SI{80.30}{\mega\Hz} yields $f_{\rm ERROR}$ = \SI{0}{\kilo\Hz} and  $f_{\rm ERROR}$ = \SI{500}{\kilo\Hz}, respectively.

After amplification, $f_{\rm ERROR}$ is converted to a voltage via a frequency-to-voltage converter which feeds into a Stanford Research Systems SIM 960 Analog PID controller. An internal set point on the PID controller is set to a voltage corresponding to $f_{\rm ERROR}$ of \SI{200}{\kilo\Hz}, which maps to $f_{\rm AOFS}$ of \SI{80}{\mega\Hz}. The PID controller then alters the pump power, accordingly, which in turn affects the center wavelength of the beam. 

\subsection{\label{sec:performance}Performance}
The proceeding sections will present short- and long-term performance characterization of the system as described above with an emphasis on long-term stabilization performance.

\subsubsection{\label{sec:shortTermPerf} Performance in the short term}
\begin{figure*} [htbp]
\centering
    \includegraphics[height=6.cm]{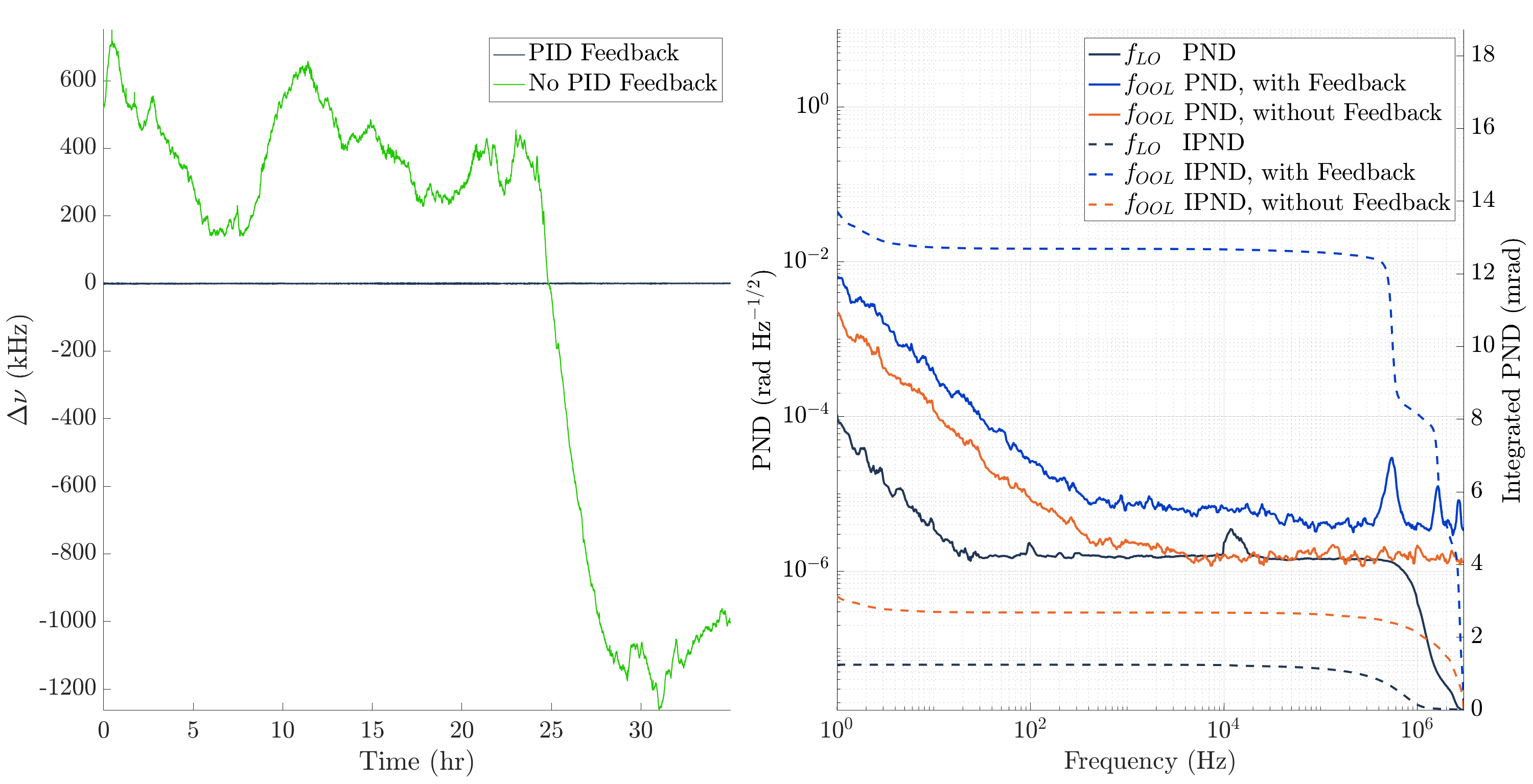}
    \caption{The plot on the left is the variation in $f_{\rm AOFS}$ with FB and without FB over a 35 hour run. The plot on the right is the phase noise and integrated phase noise of LO, OOL with FB, and OOL without FB. These plots show how FB system affects performance of experiment in regards to added phase noise in short-term and  stabilization of $f_{\rm AOFS}$ in long-term. On the left, drifts in $f_{\rm AOFS}$ are shown for two signals: one with exclusively feed-forward (in green labeled No PID Feedback) and the other with both feed-forward and feedback (in dark blue labeled PID Feedback). On the right, three signals are shown: local oscillator frequency (grey), OOL frequency with FB (blue), and OOL frequency without FB (orange). For each of these signals both PND and IPND are shown, solid and dashed lines, respectively.}
    \label{fig:shortTerm}
\end{figure*}

Without the FB loop, the integrated phase noise density (IPND) for $f_{\rm OOL}$ is \SI{3.13}{\milli\radian} (\SI{1}{\Hz} to \SI{3}{\mega\Hz}), corresponding to an rms jitter of \SI{2.57}{\atto\s}. With the FB system in place, the IPND increases to about \SI{13.72}{\milli\radian}, corresponding to an rms jitter of \SI{11.29}{\atto\s} (see second half of Table~\ref{tab:mokuData}). While this increase is significant, the addition of the FB system is not detrimental to the short-term phase noise. Figure~\ref{fig:shortTerm} (on the right) shows the PND and IPND of the system with and without the FB loop in place along with the PND and IPND for the local oscillator.

Notwithstanding such low noise performance, the system without the FB loop is unable to stabilize the CEP beyond around thirty minutes. Figure~\ref{fig:shortTerm} (on the left) compares the drifts in $f_{\rm AOFS}$ over a thirty-five hour run for an exclusively feed-forward system (labeled in figure as No PID Feedback) to our system with both feed-forward and feedback (labeled in figure as PID Feedback). Without FB, $f_{\rm AOFS}$ drifts by hundreds of kHz, whereas the FB constrains it to less than tens of kHz, which is comfortably within the bandwidth of the AOFS.

From these results it is clear that the addition of the FB system to the FF system does not impose a significant detriment to the phase noise in exchange for constraining frequency drifts in $f_{\rm AOFS}$ that otherwise would be unwieldy over practically long durations. 

\subsubsection{\label{sec:longTermPerf} Performance in the long term}
\begin{figure*} [htbp]
    \centering
    \includegraphics[height=7.2cm]{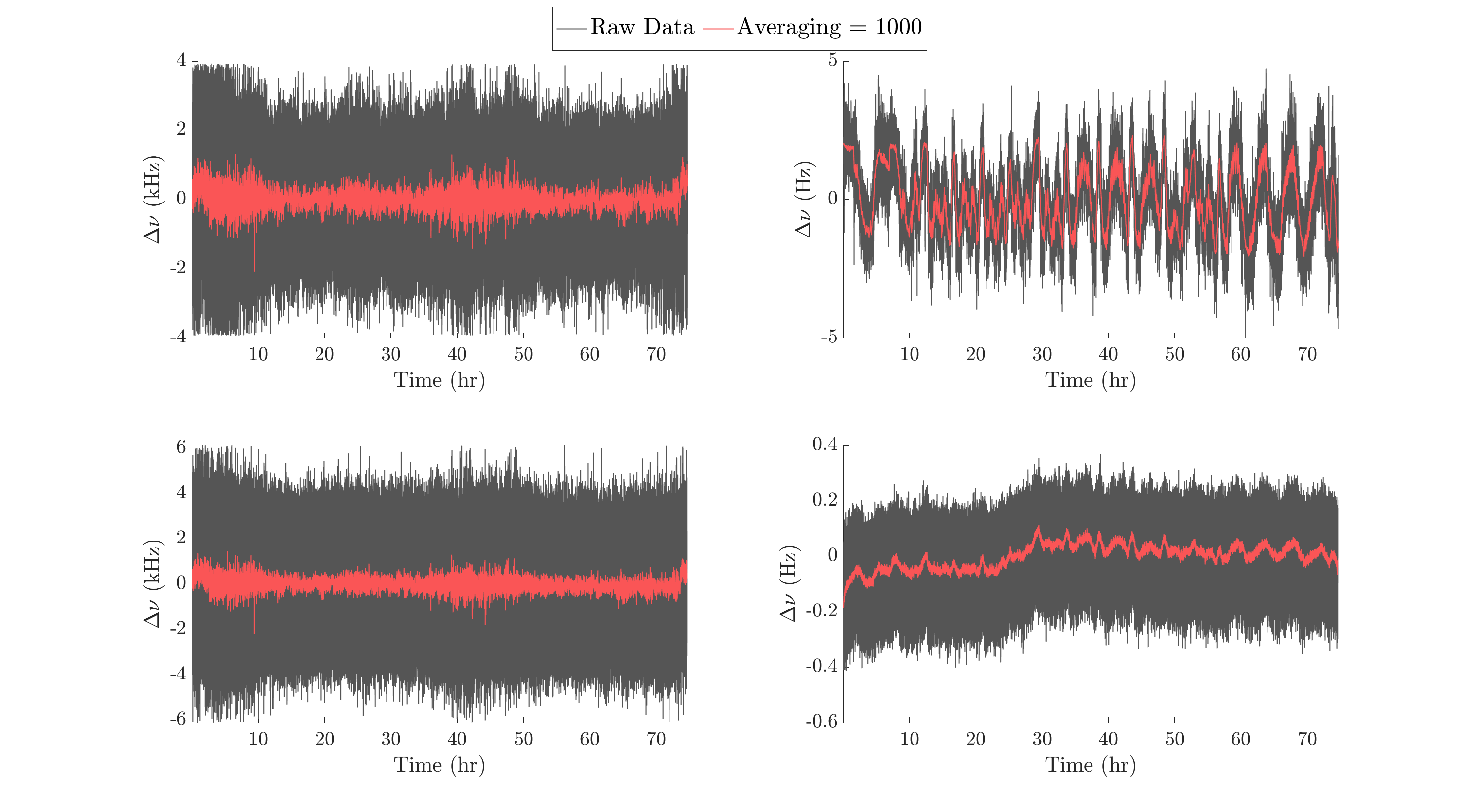}
    \caption{\label{fig:mokuRaw} Four plots show frequency drifts of $f_{\rm AOFS}$ (top left), $f_{\rm OOL}$ (top right), $f_{\rm CEO}$ (bottom left), and $f_{\rm LO}$ (bottom right) over the 75 hour run. For each, the raw signal (in grey) and averaged signal (in red) are shown.}
\end{figure*}

Using the FB system described in section~\ref{sec:FB}, over 75 hours of stabilization was recorded. Figure~\ref{fig:mokuRaw} shows the results from $f_{\rm AOFS}$, $f_{\rm OOL}$, $f_{\rm CEO}$, and $f_{\rm LO}$, all with the raw data and with a moving average of 1000. The moving average reveals slower trends in the data.

Both $f_{\rm LO}$ and $f_{\rm OOL}$ describe drifts in the local oscillator frequency. The drifts are on the order of hundreds of \SI{}{\milli\Hz} to several \SI{}{\Hz}, respectively. Since $f_{\rm AOFS}$ is the combination of $f_{\rm CEO}$ and $f_{\rm LO}$ and since $f_{\rm LO}$ has very small drifts, the drifts in $f_{\rm AOFS}$ are dominated by drifts in $f_{\rm CEO}$. These drifts are contained to less than tens of \SI{}{\kilo\Hz} with an rms jitter of \SI{0.27}{\kilo\Hz} (see Table~\ref{tab:mokuData} for rms frequency jitter for all signals). Furthermore, for this run, $f_{\rm CEO}$ was directly collected from within the IL RD conditioning block, and its drifts closely match $f_{\rm AOFS}$ drift behavior.   

\begin{figure*} [htbp]
\centering
    \includegraphics[height=6.8cm]{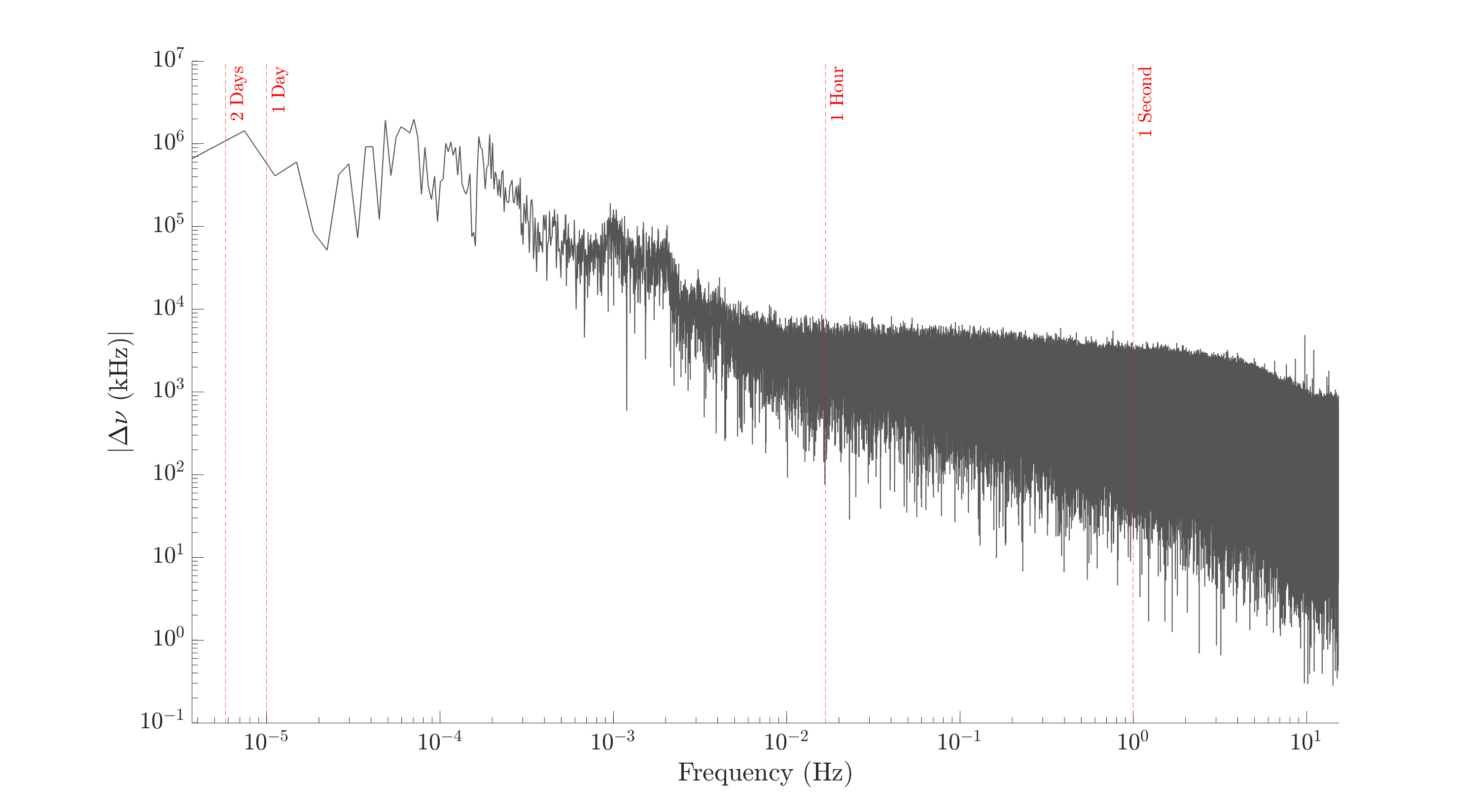}
    \caption{\label{fig:fft_aofs} Plot shows absolute value of frequency drifts of $f_{\rm AOFS}$ versus frequency. Time labels corresponding to specific frequencies are shown for 1 second, 1 hour, 1 day, and 2 days in order to contextualize noise source time-scales.}
\end{figure*}

Figure~\ref{fig:fft_aofs} shows the $f_{\rm AOFS}$ signal in the frequency domain, where low frequency noise is dominant. Aside from 1/f noise, environmental factors such as temperature, humidity, and pressure can cause fluctuations in the half-hour to few hour range. In particular, there are large drifts in the absolute value of $f_{\rm AOFS}$ for frequencies between \SI{1}{\milli\Hz} to \SI{0.1}{\milli\Hz}, corresponding to time periods of a half-day to a day that match with where the largest amplitudes in the drift of the environmental variables are. 

\begin{figure*} [htbp]
    \centering
    \includegraphics[height=6.8cm]{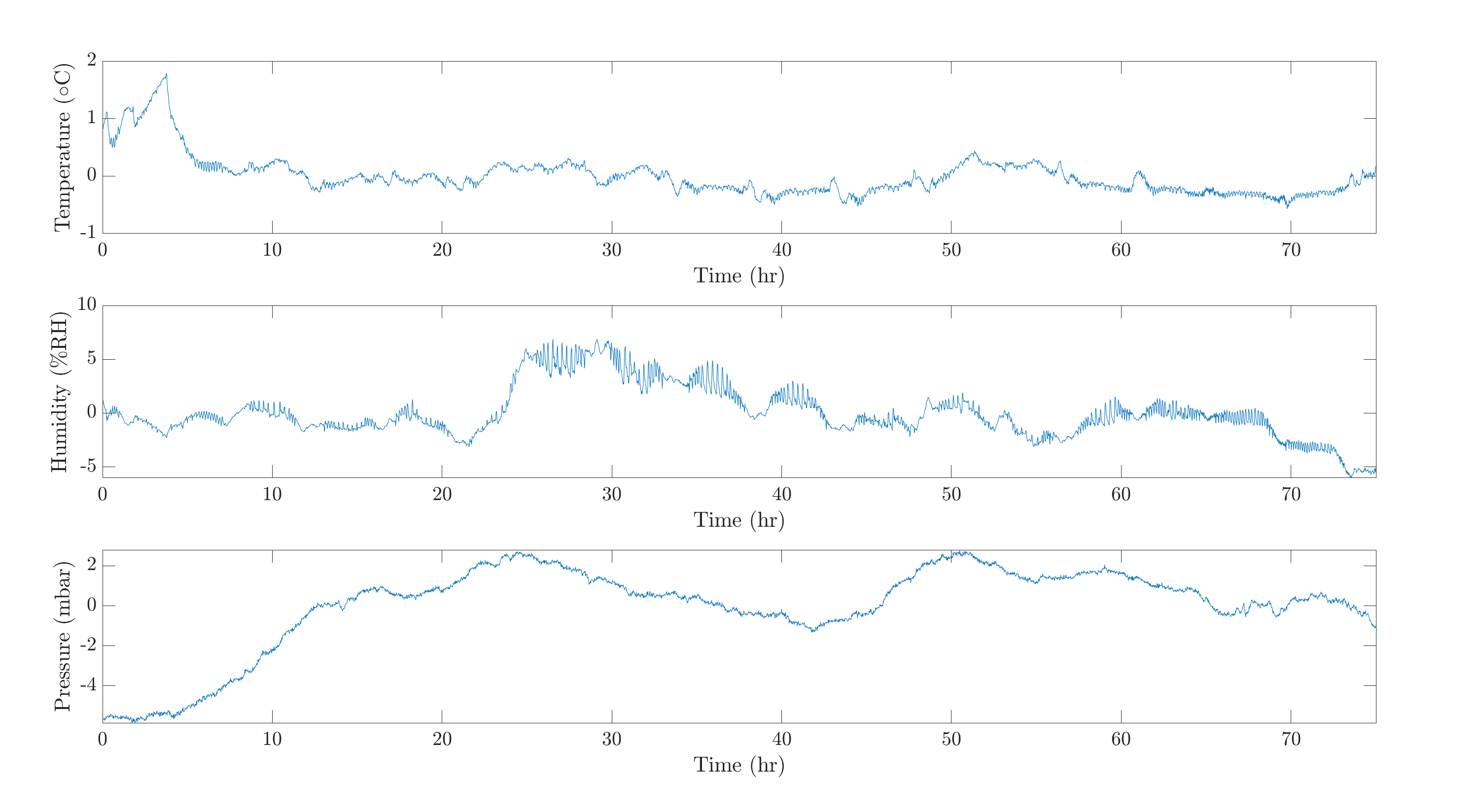}
    \caption{\label{fig:environment} Three plots show fluctuation around the mean for temperature (top), humidity (middle), and pressure (bottom) over the 75 hour run.}
\end{figure*}

Figure~\ref{fig:environment} shows the variation of these variables around their means, collected over 75 hours uninterrupted. Within the first few hours of the first day, temperature drops by nearly 2\degree C. Humidity has a change of around 12\% RH over the second 24-hour period, and pressure has a peak-to-peak change of nearly \SI{6}{\milli\bar}.  In these measurements, the laboratory was not temperature- or humidity-stabilized to high precision optical laboratory standards. Additionally, weather conditions, including relatively heavy rain within the first twenty-four hours, affected humidity and pressure.

\begin{table*}[htpb]
 \centering\caption{\label{tab:mokuData} Left half shows time lag (in hours) for the maximum correlation values from cross-covariance analysis performed for each measured frequency signal against each environmental variable along with the correlation value in square brackets. Right half shows signal stability characterization, specifically with comparisons of IPND and timing jitter for FB versus no FB.}
 \begin{adjustbox}{max width=\textwidth}
 
    \begin{tabular}{|c|c||c||c|||c|c|c||c|c|}
    \hline
     & Temperature & Humidity & Pressure & \multicolumn{3}{c||}{FF + FB} & \multicolumn{2}{c|}{FF only} \\
    \cline{1-9}
    \rule{0pt}{3ex}  &  &   &  & RMS Freq & IPND & Timing & IPN & Timing\\
    & Time Lag (hrs) [corr] & Time Lag (hrs) [corr] & Time Lag (hrs) [corr] & Jitter (Hz) & (mrad)& Jitter (as)&(mrad)& Jitter (as)\\ 
    \cline{2-9}
    \rule{0pt}{3ex} $f_{\rm AOFS}$ & 0.00 [0.06]  & 3.85 [-0.03]  & 0.15 [-.05] & 0.25 k& -  &- & - &  -\\ 
    \rule{0pt}{2ex}$f_{\rm OOL}$  & 3.79 [0.19]   & 25.48 [-0.15]  & 0.00 [-0.25] & 0.04 & 13.72 & 11.29 & 3.13& 2.57\\
    \rule{0pt}{2ex}$f_{\rm CEO}$  & 0.00 [0.06]  & 36.74 [-0.03]   & 0.16 [-0.04] & 0.27 k& - & - & -   & -\\
    \rule{0pt}{2ex}$f_{\rm LO}$ & 0.00 [-0.31]   & 3.99 [0.22]  & 13.27 [0.29] & 1.11 & 1.25 &  1.03& 1.25 & 1.03 \\
    \hline
    \end{tabular}
\end{adjustbox}
\end{table*}

Cross-covariance analysis was performed on each environmental variable with each frequency signal. The data were normalized before performing the correlation analysis and the cross-covariance values were normalized to values between $\pm 1$. The first half of Table~\ref{tab:mokuData} shows the time lag values for the maximum correlation values from these analyses along with the correlation values within brackets. These time lag values correspond to when effects from the environment are correlated to changes in the data. For instance, the maximum correlation between humidity and $f_{\rm AOFS}$ occurs roughly after a 3.85 hour delay between when the change in temperature potentially manifests as a change in the $f_{\rm AOFS}$ data. $f_{\rm AOFS}$ and $f_{\rm CEO}$ have similar correlation values with temperature and pressure. The time lag of these correlations are on the order of minutes. $f_{\rm LO}$ with temperature and $f_{\rm LO}$ and $f_{\rm OOL}$ with pressure have the strongest correlations relative to the whole set. More generally, the correlations between $f_{\rm LO}$ and $f_{\rm OOL}$ with the environmental data are stronger than those of $f_{\rm AOFS}$ and $f_{\rm CEO}$ with the environmental data. 

We believe the step in the $f_{\rm LO}$ around the 30 hour mark is likely caused by the drop in humidity around the same time. Additionally, we suspect that the oscillatory behavior in $f_{\rm OOL}$ could be coupling in from RF noise or signals in the FB system imprinted on the optical comb driving the AOFS.

\section{\label{sec:conclusions}Conclusions}
Controlling CEP for mode-locked lasers has continued to rise in importance with a vast array of experiments requiring a stabilized CEP over practically long duration. A mix of feed-forward and feedback techniques provides one such avenue for accomplishing this stabilization. 

The system presented in this report combines a FF portion, consisting of an RF signal driving an AOFS which modulates the beam downstream, and a slow FB portion, which uses the deviation of $f_{\rm AOFS}$ as the error signal for a PID control loop modulating the pump power to the cavity. The FF technique, alone, selects $f_{\rm CEO}$, keeping it stable for short periods of time until environmental effects and other noise alter $f_{\rm CEO}$ to the point that the AOFS cannot recover stability. The FB section, thus, keeps $f_{\rm AOFS}$ at the proper frequency by altering the beam's central wavelength, allowing the FF system to remain locked. From the OOL measurements, implementing the FB with the FF added about \SI{10}{\milli\radian} to the integrated phase noise of the FF system alone. To investigate potential drift sources, we collected data and studied the correlation of environmental factors (local temperature, humidity, and pressure) and changes in the performance of the system.  

Using these combined methods, we were able to show stabilization for over 75 hours with \SI{13.72}{\milli\radian} integrated phase noise to be used in real-world applications under unfavorable environmental conditions. The 75-hour run was interrupted only for reporting purposes and our data indicates that the system would have remained locked indefinitely. 

\section*{Acknowledgements}
Use of the Linac Coherent Light Source, SLAC National Accelerator Laboratory, is supported by the U.S. Department of Energy, Office of Science, Office of Basic Energy Sciences, under contract no. DE-AC02-76SF00515

\section*{Disclosures}
The authors declare no conflicts of interest.

\nocite{*}
\bibliography{main}

\end{document}